\documentclass{acm_proc_article-sp}

\usepackage{graphicx}
\usepackage{enumitem}

\makeatletter
\def\@copyrightspace{\relax}
\makeatother

\title{Undefined By Data: A Survey of Big Data Definitions}
\begin{document}

\numberofauthors{1} 

\author{
\alignauthor Jonathan Stuart Ward and Adam Barker\\
	\affaddr{School of Computer Science} \\
	\affaddr{University of St Andrews, UK}\\
	\email{\{jonthan.stuart.ward, adam.barker\}@st-andrews.ac.uk}
}

\maketitle

\begin{abstract}
The term big data has become ubiquitous. Owing to a shared origin between academia, industry and the media there is no single unified definition, and various stakeholders provide diverse and often contradictory definitions. The lack of 
a consistent definition introduces ambiguity and hampers discourse relating to big data. This
short paper attempts to collate the various definitions which have gained some degree of traction
and to furnish a clear and concise definition of an otherwise ambiguous term.
\end{abstract}

\section{Big Data}
Since 2011 interest in an area known as big data has increased
exponentially~\cite{Google2013}. Unlike the vast majority of computer science
research, big data has received significant public and media interest.
Headlines such as ``Big data: the greater good or invasion of
privacy?''~\cite{Chatterjee2013} and ``Big data is opening doors, but maybe too
many''~\cite{Lohr} speak volumes as to the common perception of big data. From
the outset it is clear that big data is intertwined with considerable technical
and socio-technical issues but an exact definition is unclear. Early literature
using the term has come from numerous fields. This shared provenance has led to
multiple, ambiguous and often contradictory definitions. In order to further
research goals and eliminate ambiguity, a concrete definition is necessary. 

Anecdotally big data is predominantly associated with two ideas: data storage
and data analysis.  Despite the sudden interest in big data, these concepts
are far from new and have long lineages. This, therefore, raises
the question as to how big data is notably different from conventional data processing 
techniques. For rudimentary insight as to the answer to this question one
need look no further than the term big data. ``Big'' implies significance, complexity
and challenge. Unfortunately the term ``big'' also invites quantification and therein lies
the difficulty in furnishing a definition. 

Amongst the most cited definitions is that included in a Meta
(now Gartner) report from 2001~\cite{douglas20013d}.  The Gartner report makes
no mention of the phrase ``big data'' and predates the current trend. However,
the report has since been co-opted as a key definition. Gartner proposed a
three fold definition encompassing the ``three Vs'': Volume, Velocity, Variety.
This is a definition routed in magnitude. The report remarks upon the
increasing size of data, the increasing rate at which it is produced and the
increasing range of formats and representations employed. As is common
throughout big data literature, the evidence presented in the Gartner
definition is entirely anecdotal. No numerical quantification of big data is
afforded. This definition has since been reiterated by NIST~\cite{NIST2013} and Gartner in 2012~\cite{beyer2012importance} expanded upon by IBM~\cite{IBM-bigdata} and others to include
a fourth V: Veracity. Veracity includes questions of trust and uncertainty with
regards to data and the outcome of analysis of that data.

Oracle avoids employing any Vs in offering a definition. Instead Oracle~\cite{dijcks2012oracle} contends that big 
data is the derivation of value from traditional relational database driven business decision making,
augmented with new sources of unstructured data. Such new sources include blogs, social media, sensor
networks, image data and other forms of data which vary in size, structure, format and other factors. Oracle,
therefore asserts a definition which is one of inclusion. They assert that big data is the inclusion of additional
data sources to augment existing operations. Notably, and perhaps unsurprisingly,
the Oracle definition is focused upon infrastructure. Unlike those offered by others, Oracle places
emphasis upon a set of technologies including: NoSQL, Hadoop, HDFS, R and relational databases. In doing so
they present both a definition of big data and a solution to big data. While this definition is somewhat more 
easily applied than others it similarly lacks quantification.  Under the Oracle definition it is not clear as
to exactly when the term big data becomes applicable it rather provides a means to ``know it when you see it''.

Intel is one of the few organisations to provide concrete figures in their literature. Intel links big data
to organisations ``generating a median of 300 terabytes (TB) of data weekly''~\cite{intel-bigdata}. Rather than providing
a definition as per the aforementioned organisations, Intel describes big data through quantifying the experiences
of its business partners. Intel suggests that the organisations which were surveyed deal extensively with unstructured data and place an emphasis on performing
analytics over their data which is produced at a rate of up to 500 TB per week. Intel asserts that the most common data type involved in analytics is business transactions
stored in relational databases (consistent with Oracle's definition), followed by documents, email, sensor data, blogs 
and social media. 

Microsoft provides a notably succinct definition: ``Big data is the term
increasingly used to describe the process of applying serious computing power - 
the latest in machine learning and artificial intelligence - to seriously
massive and often highly complex sets of information''~\cite{Microsoft2013}. This definition states in 
no uncertain terms that big data requires the application of significant compute power. This
is alluded to in previous definitions but not outright stated. Furthermore this definition
introduces two technologies: machine learning and artificial intelligence which have 
been overlooked by previous definitions. This, therefore, introduces the concept of 
there being a set of related technologies which are form crucial parts of a definition.

A definition, or at least an indication of related technologies can be obtained through an investigation of related terms. Google
Trends provides the following terms in relation to big data~\cite{Google2013}, from most to least frequent: data analytics, Hadoop, NoSQL, Google, IBM, 
and Oracle. From these terms a number of trends are evident. Firstly, that big data is intrinsically related to
data analytics and the discovery of meaning from data. Secondly, it is clear that there are a number of related technologies as alluded to 
by the Microsoft definition, namely NoSQL and Apache Hadoop. Finally it is
evident that there are a number of organisations, specifically industrial
organisations which are linked with big data.

As suggested by Google Trends, there are a set of technologies which are frequently suggested as being 
involved in big data. NoSQL stores including Amazon Dynamo, Cassandra, CouchDB,
MongoDB et al play a critical role in storing large volumes of unstructured and highly variable data.
Related to the use of NoSQL data stores there is a range of analysis tools and
methods including MapReduce, text mining, NLP, statistical programming, machine learning and
information visualisation. The application of one of these technologies alone is not sufficient to merit the use of
the term big data. Rather, trends suggest that it is the combination of a number of technologies and 
the use of significant data sets that merit the term. These trends suggest big data as a technical movement
which incorporates ideas, new and old and unlike other definitions provides little commentary as to social and business implications.

While the previously mentioned definitions rely upon a combination size,
complexity and technology, a less common definition relies purely upon
complexity. The Method for an Integrated Knowledge
Environment (MIKE2.0) project, frequently cited in the open source community, introduces a potentially contradictory idea:
``Big Data can be very small and not all large datasets are big''~\cite{mike}. This is an
argument in favour of complexity and not size as the dominant factor.  The MIKE
project argues that it is a high degree of permutations and interactions within
a dataset which defines big data. 

The idea expressed latterly in the MIKE2.0 definition; that big data is not
easily handled by conventional tools is a common anecdotal definition. This
idea is supported the NIST definition which states that big data is data which:
``exceed(s) the capacity or capability of current or conventional methods and
systems''~\cite{NIST2013}.  Given the constantly advancing nature of computer
science this definition is not as valuable as it may initially appear. The
assertion that big data is data that challenges current paradigms and practices
is nothing new. This definition suggests that data is ``big'' relative to the
current standard of computation. The application of additional computation or
indeed the advancing of the status quo promises to shrink big data.  This
definition can only serve as a set of continually moving goalposts and suggests
that big data has always existed, and always will.

Despite the range and differences existing within each of the aforementioned definitions
there are some points of similarity. Notably all definitions make at least one of
the following assertions:
\begin{description}[noitemsep,nolistsep]
	\item[Size:] the volume of the datasets is a critical factor.
	\item[Complexity:] the structure, behaviour and permutations of the datasets is a critical factor.
	\item[Technologies:] the tools and techniques which are used to process a sizable or complex dataset is 
		a critical factor.
\end{description}
The definitions surveyed here all encompass at least one of these factors, most encompass two. An extrapolation
of these factors would therefore postulate the following: \textit{Big data is a term describing the storage and analysis of large and or complex data sets using a series
of techniques including, but not limited to: NoSQL, MapReduce and machine learning.}
\bibliographystyle{abbrv}
\bibliography{bdl.bib}
\end{document}